# Virtual Screening of Pharmaceutical Compounds with hERG Inhibitory Activity (Cardiotoxicity) using Ensemble Learning

Aditya Sarkar[1], Arnav Bhavsar[1]

[1]School of Computing and Electrical Engineering, Indian Institute of Technology Mandi, Kamand, Mandi, HP, India

**Abstract**

In silico prediction of cardiotoxicity with high sensitivity and specificity for potential drug molecules can be of immense value. Hence, building machine learning classification models, based on some features extracted from the molecular structure of drugs, which are capable of efficiently predicting cardiotoxicity is critical. In this paper, we consider the application of various machine learning approaches, and then propose an ensemble classifier for the prediction of molecular activity on a Drug Discovery Hackathon (DDH) (1st reference) dataset. We have used only 2-D descriptors of SMILE notations for our prediction. Our ensemble classification uses 5 classifiers (2 Random Forest Classifiers, 2 Support Vector Machines and a Dense Neural Network) and uses Max-Voting technique and Weighted-Average technique for final decision.

**Introduction**

It is well known that drug discovery is complex, long drawn, and requires interdisciplinary expertise to discover new molecules. Drug safety is an important issue in the process of drug discovery. Failure in clinical trials in the 2000s was majorly due to efficacy and safety (approx 30%) (Kola, I. and Landis, J., 2004). One important aspect of drug safety is drug toxicity. Frequently observed toxicities are cardiotoxicity, hepatotoxicity, genotoxicity, and phototoxicity (Keiji Ogura, 2019). Toxicological screening is very important for the development of new drugs and for the extension of the therapeutic potential of existing molecules. The US Food and Drug Administration (FDA) states that it is essential to screen new molecules for pharmacological activity and toxicity potential in animals (21 CFR Part 314). The toxic effects of chemicals, food substances, pharmaceuticals, etc., have attained great significance in the 21st century (Parasuraman, S, 2011). The h-ERG (human Ether-`a-go-go-Related Gene) is a gene that codes for a protein known as Kv11:1, the alpha subunit of a potassium ion channel ("hERG safety",2018). The h-ERG potassium channels (Snyders, 1999) are essential for normal electrical activity in the heart. When this channel's conductivity of electric current is inhibited by some action of drugs, it can lead to a fatal disorder called Long QT Syndrome. It is found that many drugs have the h-ERG inhibitory activity which can prolong the QT and thereby resulting in irregularity of the heartbeat called Torsades de Pointes (Keiji Ogura et al, 2019). As a result, many drugs, that are inhibiting the h-ERG channel's conductivity, have been withdrawn from the markets. Hence, it is regarded as a major anti-target for drug discovery. Over the years, many works have been done to achieve the goal of classifying compounds having h-ERG inhibitory activity. In early drug discovery stages such as screening h-ERG inhibitory activity, performing costly and time-consuming assays is difficult. Hence, developing an in-silico model to predict hERG inhibition is very useful. (Keiji Ogura et al, 2019) Machine Learning techniques can be used for classifying if any compound is having the inhibitory activity or not. Machine Learning models can learn features that can classify any compound on the basis of h-ERG inhibition activity. There have been some recent classification models reported for h-ERG inhibition which have used Neural Networks (B. Mehlig, 2019), Random Forest Classifiers (RF) (Leo Breiman, 2001) and Support Vector Machines (SVM) (Yongli Zhang et al, 2012). Below we provide a brief review of these.

**Related Work**

Czodrowski (Czodrowski, 2013) constructed RF models using descriptors calculated by Rdkit, based on 3,721 compounds measured in a binding assay and 765 compounds measured in a functional assay collected from ChEMBL. The prediction models were constructed separately from each data set and showed prediction accuracies of 79.7%–80.1% and 69.2%–90.7%, respectively. Wang et al. (Wang, S. et al, 2012) developed hERG classification models using naive Bayesian classification and recursive partitioning based on molecular properties and the ECFP 8 fingerprints, and recorded 85% accuracy (Wang, S. et al, 2016). Schyman et al. (Schyman. P, 2016) combined 3D (David C. Kombo et al,2012) similarity conformation and 2D similarity ensemble approach, and achieved 69% sensitivity and 95% specificity on an independent external data set. Recently, Keiji Ogura, Tomohiro Sato, Hitomi Yuki and Teruki Honma (Keiji Ogura et al, 2019) used Support Vector Machines (SVMs) on an integrated h-ERG database having more than 291,000 structurally diverse compounds. They achieved kappa statistics of 0.733 and accuracy of 98.4%. Supplementary Table S1 provides a summary of various datasets used across existing works. They have made the dataset publicly available for research purposes.

**Contributions to this work**

Most of the works (Wang, S. et al, 2012), (Wang, S. et al, 2016), (Schyman. P, 2016), (Keiji Ogura et al, 2019) which we have reviewed have either used descriptors (2-D, 3-D and 4-D) or fingerprints. On the other hand, unlike the above, we have used only 2-D descriptors for our classification model. 2-D descriptors deal with the molecular topology of the compounds i.e. topological indices and fragment counts. 2D-Descriptors incorporate precious chemical information like size, degree of branching, flexibility etc. Generating 2D descriptors of the SMILES compounds usually takes less time than 3D descriptors. Even with just 2D descriptors, we demonstrate that the proposed ensemble model achieves a very good performance. In this study, we have developed an ensemble model having two Random Forest Classifiers, two Support Vector Machines and one Dense Neural Network which achieved an AUC score (Area Under the ROC Curve) of 0.96 and Cohen's Kappa of 0.9195. Most of the existing models have used Support Vector Machines, Random Forest Classifiers and Naive Bayesian Classifiers for prediction. However, in addition to these models, we have also used Deep Neural Networks and two different Ensemble classifiers for the task. We have found that the Deep Neural Networks and the Ensemble classifier yield the highest performance. We have also worked with data augmentation for our class-imbalanced dataset. We have used SMOTE (Synthetic Minority Oversampling Technique) (N. V. Chawla, et al, 2011) for augmenting data. Data augmentation is a very useful procedure because the data that it generates is almost similar to the training data. For some diseases, the drugs available can be quite less, and doing prediction with less data points can lead to overfitting. Data Augmentation can prove useful by not only creating new data but may also help in understanding the underlying distribution of each property (descriptors or fingerprints) of drugs. Unlike most existing works, we also suggest an automatic approach based on information gain/entropy to shortlist (or select) features from a larger set. To our knowledge, the only exception among the existing methods, which are considered such a selection is work by (Keiji Ogura et al, 2019) which involves the NSGA-II (Non-dominated Sorting Genetic Algorithm-II) for descriptor selection. Finally, towards the end of the paper, we provide a consolidated summary of the various works in this domain, the datasets used, the feature descriptors and methods employed, and their performance across several metrics. We note that although this work is not analyzing imaging data, it involves core machine learning on an important problem in biology. Considering that this is a relatively recent application domain, such an overview provides a good perspective of the trade-offs of the approaches and paves the way for more standardized benchmarking and extensions in this area.

**Dataset and Descriptors**

In this work, we have used a dataset provided in one of the competitions of the Drug Discovery Hackathon, organized by the Govt. of India; The dataset has been made by Dr Kunal Roy, professor from Department of Pharmaceutical Technology, Jadavpur University. http://sites.google.com/site/kunalroyindia/home. It contains the SMILE notations of 8227 pharmaceutical compounds, along with their h-ERG inhibitory activity label (i.e. blocker or non-blocker). Out of these, 6878 compounds were blockers and 1349 were non-blockers. We have used Mordred (Moriwaki H, et al, 2018) Python module to decrypt the SMILE (Anderson, E. et al, 1987) notations to 2-D descriptors. As a result, we got 1613 features for each compound. Figure 1 shows a snippet of the Pandas view of the dataset. It shows the IDs of the pharmaceutical compounds, the class to which it belongs i.e. Blocker or Non- Blocker (Blocker means that the compound possesses the h-ERG inhibitory activity and Non-Blocker means that it does not possess the h-ERG inhibitory activity) and few 2D descriptors. The list of descriptors along with their descriptions can be found at Mordred documentation. Initially, we have extracted all the 2-D descriptors that Mordred has offered. In later sections, we have described a method for further selecting features out of these.

**Method**

We have first worked on data cleaning, then augmented our data because it was having an imbalanced class problem, selected important features using an Entropy/Information Gain based method, and finally performed the classification using two ensembles - Max-Voting Ensemble and Weighted-Average Ensemble using two variants of Random Forest Classifiers, two variants of Support Vector Machines and one Deep Neural Network. We have divided this section into 5 subsections - Data Cleaning, SMOTE Application, Feature Selection and Base Models and Ensemble-Learning.

**Data Cleaning**

Data Cleaning is an important part of analyzing this data. This is done so as to eliminate outliers present in the data. These outliers are due to miscalculations made by the Mordred python module. Since the range of each column is different, to normalize them, Z-score/ Standard Score is used. Z-score or standard score of a particular column is defined as the number of standard deviations by which the value of a datapoint value is above or below the mean value of data points present in the column. Raw scores above the mean have positive Z-Scores, while those below the mean have negative Z-scores (Spiegel, Murray R.; Stephens, Larry J, 2008). Mathematically, it is defined as - Z-Score = X-μ/$\sigma$ where x is the sample datapoint, μ is mean of all the data points in the sample column and $\sigma$ is standard deviation in the sample column. We have used Z-scores for finding the outliers and finally replaced the outliers with the mean of column of the Dataframe, to which it belongs. The Dataframe is two-dimensional, size-mutable, potentially heterogeneous tabular data. This can be interpreted as removing the outlier samples, and augmenting the rest of the samples, with a mean estimate. We have considered a datapoint, an outlier when |Z-Score(datapoint)|>3 (PeruriVenkataAnusha et. al.). A section of the dataset view obtained after cleaning is shown in Figure 1. It shows the class to which the pharmaceutical compound belongs to i.e. Blocker or Non-Blocker and few 2D descriptors. However, unlike Figure 1, this contains a Z-score. Also, we have removed those features which were containing NaN values. Hence, our number of features decreased from 1613 to 1375.

**SMOTE for Data Augmentation**

As mentioned earlier, our dataset has 6878 Blocker compounds and 1349 non-Blocker compounds. Hence, it is imbalanced and can lead to high bias. To tackle this problem, we made use of the popular data augmentation method SMOTE (Synthetic Minority Oversampling Technique). SMOTE (N. V. Chawla, 2011) can be used to create synthetic examples for the minority class. It works by first choosing a random example from the minority class and then k of the nearest neighbors for that example are found. A randomly selected neighbor is interpolated between the two examples in feature space. We chose SMOTE considering its popularity, with which we were also getting an improvement in our results. However, one can also consider other data augmentation methods. We have used Imblearn Python Module (Guillaume Lemaitre et al, 2017) for applying SMOTE. After SMOTE, we achieved a total size of 13756 data points, including 6878 Blocker compounds and 6878 Non-Blocker compounds.

**Feature extraction**

Feature selection is an important part of the model. Considering we have numerous 2-D features, it is important to consider the ones which can contribute to the task of the classification. Gini-index and Entropy are used as criteria for calculating information gain. Decision tree algorithms use information gain to split a node. Entropy and Gini are used for measuring impurity of a node. Nodes having multiple classes are considered impure and nodes having a single class are considered pure. In this project, we have used Entropy as our impurity measure. While training a tree, we can compute how much each feature decreases the impurity. The more a feature decreases the impurity, the more important that feature is. In random forests, the impurity decrease from each feature can be averaged across trees to determine the final importance of the variable. For this we have used the features selected by Random Forests with measure of impurity as Entropy. Initially, we started with 1375 features for each compound. After applying feature selection, we had 592 features, which is a significant reduction.

**Base models and ensemble learning**

We have used 5 base models in our ensemble - 2 Random Forest Classifiers, 2 Support Vector Machines and 1 Dense Neural Network. These well-known methods are discussed briefly in the following subsections, and the corresponding parameters are provided in Table 2. We have used the Scikit-Learn Python module for Random Forest Classifiers and Support Vector Machines. For implementing Dense Neural Networks, we have used Tensorflow 2 and Keras.

The random forest classifier is essentially an ensemble of decision tree-based classifiers. It operates by constructing a number of decision trees during its training and outputs a class decision that is the mode of the class estimates of each decision tree. It is based on the principle of bagging to mitigate the bias-variance trade-off in decision tree-based classifiers.

SVM is a popular maximum-margin classification framework boasting advantages of good generalization and non-linear classification via the use of kernel functions. Support Vector machines find a maximum margin- hyperplane that divides the data points of the classes such that the distance between the hyperplane and the nearest point of either class to this hyperplane is maximized.

Dense Neural Networks have gained popularity as contemporary classifiers due to their ability to learn highly non-linear classification models, given enough data. A neural network is a network of neurons that can well approximate a highly nonlinear boundary between the classes, given enough data.

For our Ensemble, we have used Max-Voting and Weighted-Average with our 5 base models described above. In max-voting, each base model makes a prediction and votes for each sample. Only the sample class with the highest votes is included in the final predictive class. In weighted average, we have placed weights on predictions of each of the base models for the final prediction. The weights we have placed are 0.75 for DNN, 0.1 for RF-1, 0.07 for RF-2, 0.05 for SV-1 and 0.03 for SV-2. We have given more weights to that classifier among the base models, which yields a higher accuracy. Since, our Neural network is showing maximum accuracy, we have placed maximum weightage to it.

**Experiments and outcomes**

For our experiments, we have divided our dataset in the training-testing ratio of 70% and 30%. As a result, our training set contains 9629 data points and the testing set has 4127 data points. The training data is split into training and validation sets, automatically by the inbuilt models in the packages that we employ. We use testing data only for prediction. We have tested the performance of the base classifiers as well as the ensemble classifiers using several metrics, in addition to the overall accuracy that are described below:

AUC-ROC score computes Area Under the Receiver Operating Characteristic Curve (ROC) from prediction scores. The ROC curve is shown in Figure 2. We note from all ROC curves that a high True Positive is achieved at fairly low value of False positives.

Abbreviating TN for True Negative, TP for True Positive, FN for False Negative and FP for False Positive, the other metrics is defined as Sensitivity = TP/TP+FN, Specificity = TN /TN+FP, Balance Accuracy = Sensitivity+Specificity/2, Precision = TP/TP+FP.

We have also used Cohen's-Kappa (k) (J. Cohen, 1960), which determines the level of agreement between two annotators in a classification problem. $k = p_0 - p_E / 1 - p_E$ where $p_0$ is defined as the empirical probability of agreement on the label assigned to any sample and $p_E$ is defined as expected agreement when both annotators assign labels randomly.

Matthews correlation coefficient (MCC) (Matthews, B.W., 1975) is generally regarded as a balanced measure which can be used in case of data imbalance between classes.

**Results**

The accuracy and the ROC-AUC values of the base models as well as the ensemble models are provided in Supplementary Table S3. We note that among the base classifiers, the neural network model performs the best. The accuracy of the ensemble model with average weighting is similar to the DNN model. It is likely that the ensemble learning method with max voting is performing relatively lower, because of some low performing base classifiers. In Supplementary Table S4, we provide the results of the top two performing classifiers from Supplementary Table S3., for the other metrics that were defined above. We note that both the approaches yield high quality classification across all the metrics, and their performance is close to each other. While for this dataset, the DNN model performs somewhat better than the ensemble learning approach, it is important to acknowledge the high performance of the ensemble learning, as for larger datasets, it is known that the ensemble strategies can typically better mitigate the bias-variance tradeoff.

In Supplementary Table S5, we provide an overall summary of the various existing methods for the task of classification of pharmaceutical compounds based on their hERG inhibition activity. We note that while there have been a few (but not many) methods to address this task, these involve different datasets, and different features. Thus, while such a summary is not a comparison, it does provide, under one roof, a perspective on data, methods, and can help in identifying scope of improvements in this area. In the table, some cells are blank as not all metrics are provided for all methods. We note that the methods do not yield a high performance across all the metrics. Some methods employ relatively small data. The work by Ogura et al, involves the largest data, but yields a low sensitivity and kappa coefficient. Such discrepancy suggests an issue with the data imbalance. Importantly, most of the approaches use various different features in their methodology. In contrast our approach uses only 2D features and yields a high performance across all metrics. A limitation of this work is that it also involves relatively less data, which we plan to address in future.

**Conclusion and future work**

In this work we have compared various standard machine learning methods for the task of pharmaceutical compound classification based on their hERG inhibition activity. Some of the important aspects that we have considered are only 2D features, data augmentation, feature selection, and ensemble learning. The accuracy we have achieved with our model is quite high for a small dataset. The work encourages us to further explore more ensemble strategies considering DNN features, stacking, bagging etc.

**Declarations**:

**Availability of data and materials**

Dataset can be accessed using this link : https://github.com/aditya-sarkar441/h-ERG_publication

**Competing interests**

The authors declare that there are no competing interests.

**Authors contributions**

AS and AB conceived of the presented methodology. AS conducted the major analysis of the datasets and wrote the scripts for initial data preprocessing and analysis. AS and AB contributed to the writing of the manuscript. All authors discussed the text and commented on the manuscript. All authors read and approved the final manuscript.

**Figures**

| Class | ABC | ABCGG | nAcid | nBase | SpAbs_A | SpMax_A | SpDiam_A |
|---|---|---|---|---|---|---|---|
| Blocker | 24.12715927 | 17.01536288 | 0 | 1 | 40.9860129 | 2.438575691 | 4.877151381 |
| Blocker | 28.20839873 | 21.04156203 | 0 | 2 | 45.14180786 | 2.466798596 | 4.861947636 |
| Blocker | 28.52608579 | 20.57461761 | 0 | 1 | 44.6206306 | 2.474900325 | 4.949800651 |
| Blocker | 26.10634906 | 19.63229844 | 0 | 1 | 41.75702223 | 2.479655256 | 4.959310512 |
| Blocker | 29.30214225 | 21.50162648 | 0 | 1 | 45.6494774 | 2.480532387 | 4.961064773 |
| Non-Blocker | 27.58951231 | 20.53723242 | 0 | 0 | 44.07238789 | 2.480572525 | 4.96114505 |
| Non-Blocker | 26.36505935 | 20.61429083 | 0 | 1 | 43.27962018 | 2.51147146 | 5.000714285 |
| Blocker | 39.8781762 | 26.33737013 | 0 | 2 | 64.5033499 | 2.757399442 | 5.442636086 |
| Blocker | 0 | 0 | 1 | 0 | 0 | 0 | 0 |
| Non-Blocker | 11.38265818 | 9.637845481 | 0 | 0 | 19.75771955 | 2.268622815 | 4.53724563 |

| Class | ABC | ABCGG | nAcid | nBase | SpAbs_A | SpMax_A | SpDiam_A |
|---|---|---|---|---|---|---|---|
| 1.0 | 0.345196 | 0.137589 | -0.208247 | 0.131433 | 0.305502 | 0.118906 | -0.056434 |
| 0.0 | 0.901727 | 1.625585 | -0.208247 | 0.131433 | 0.631075 | 0.072039 | 0.131635 |
| 1.0 | 0.238093 | -0.022127 | -0.208247 | 0.131433 | 0.187170 | -0.242721 | -0.517803 |
| 1.0 | -0.184283 | 0.173526 | 0.000000 | -1.362078 | -0.411322 | -0.610173 | -0.355930 |
| 1.0 | -0.042343 | -0.353782 | 0.000000 | 0.131433 | -0.022795 | -0.437377 | -0.737187 |
| 1.0 | 0.386357 | 0.938283 | -0.208247 | 0.131433 | 0.374127 | 0.136865 | 0.191273 |
| 1.0 | -1.524310 | -1.389215 | -0.208247 | 0.131433 | -1.379470 | -0.141355 | -0.440691 |
| 1.0 | -0.705384 | -1.010657 | -0.208247 | 0.131433 | -0.787281 | -0.943904 | -0.736060 |
| 1.0 | 0.920544 | 1.226106 | -0.208247 | -1.362078 | 0.752153 | 0.261510 | 0.091212 |
| 0.0 | 0.598630 | 0.200595 | -0.208247 | 0.131433 | 0.513556 | -0.510563 | -0.242471 |

(a) Pandas view of Dataset          (b) Pandas view of cleaned Dataset

**Figure 1: DataFrames –** Figure 1(a) shows a snippet of the Pandas view of the dataset. It shows the IDs of the pharmaceutical compounds, the class to which it belongs i.e. Blocker or Non-Blocker (Blocker means that the compound possesses the h-ERG inhibitory activity and Non-Blocker means that it does not possess the h-ERG inhibitory activity) and few 2D descriptors. Figure 1(b) shows the dataset we have got after pre-processing and cleaning.

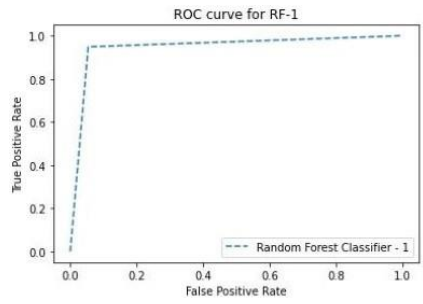

Figure 2(a)

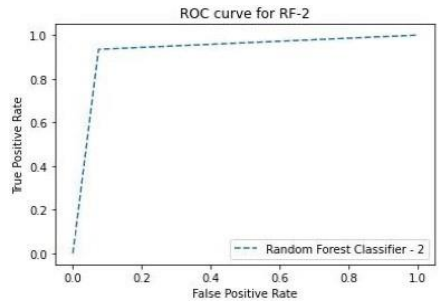

Figure 2(b)

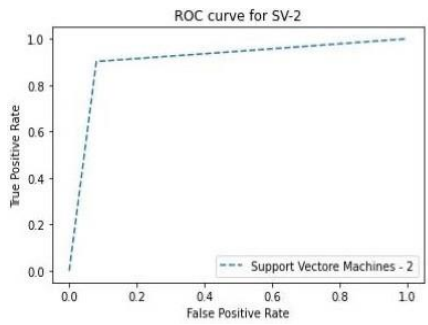

Figure 2(c)

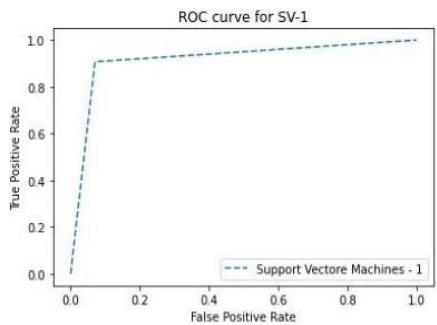

Figure 2(d)

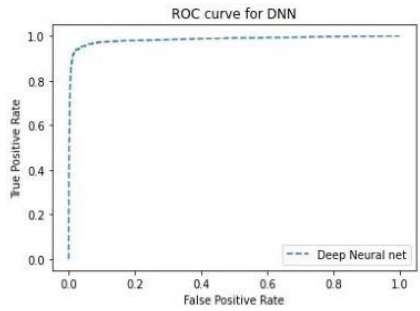

Figure 2(e)

**Figure 2: AUC-ROC plots** – The AUC-ROC plots for various models are shown here. We note from all ROC curves that a high True Positive is achieved at fairly low value of False positives.

Supplementary Tables

Table S1.

| Integrated Dataset | | |
|---|---|---|
| Database | hERG inhibitors | Inactive compounds |
| ChEMBL (version 22) | 4793 | 5275 |
| GOSTAR | 3260 | 3509 |
| NCGC | 232 | 1234 |
| hERGCentral | 4,321 | 274,536 |
| hERG integrated dataset | 9,890 | 281,329 |

Table S1 provides a summary of various datasets used across existing works. They have made the dataset publicly available for research purposes.

Table S2.

| Base Model Accuracies and ROC-AUC | |
|---|---|
| Models | Features |
| Dense Neural Networks | dense layers : 3 with L1 param $10^{-5}$, L2 param $10^{-4}$, dropout of 0.5,0.3, 0.3 resp. 1 with L1 param $10^{-4}$, L2 param $10^{-4}$, Optimiser: ADAM, Activation func: Sigmoid. |
| Random Forest Classifier-1 | trees, min no. of samples at leaf node : 2 and to split an internal node : 4, feat. selection criteria : Entropy-Based. |
| Random Forest Classifier-2 | trees, min no. of samples at leaf node : 10 and to split an internal node : 5, feat. selection criteria : Gini-Based. |
| Support Vector Machines-1 | regularization parameter : 1 |
| Support Vector Machines-2 | regularization parameter : 0.8 |

Table S2 shows the parameters we have used for the base models. Base models used are 2 Random Forests, 2 Support Vector Machines and 1 Dense Neural Network.

Table S3

| Base Model Accuracies and ROC-AUC | | |
|---|---|---|
| **Models** | **Accuracies** | **AUC** |
| Dense Neural Networks | 95.98% | 0.994 |
| Random Forest Classifier-1 | 94.3% | 0.944 |
| Random Forest Classifier-2 | 93.51% | 0.932 |
| Support Vector Machines-1 | 90.87% | 0.914 |
| Ensemble learning (Max voting) | 93.96% | 0.94 |
| Ensemble learning (Avg. weighting) | 95.97% | 0.96 |

Table S3 shows the accuracy of the base models and two ensembles. We can observe that the Dense Neural Network and the Weighted Average Ensemble Learning have the highest accuracy i.e. 95.98 % and 95.97 % respectively. However Dense Neural Network has the highest AUC score followed by Weighted Average Ensemble Learning.

Table S4

| Metrics for different Ensembles | | |
|---|---|---|
| Metrics | Weighted-Average | DNN |
| AUC | 0.960 | 0.994 |
| Sensitivity | 0.9436 | 0.975 |
| Specificity | 0.9755 | 0.983 |
| Balanced Accuracy | 0.9596 | 0.979 |
| MCC | 0.9199 | 0.956 |
| Cohen's kappa | 0.9195 | 0.956 |
| F1-score | 0.96 | 0.98 |
| Precision | 0.96 | 0.975 |
| Recall | 0.96 | 0.975 |

Table S4 shows the results of the top two performing classifiers from Table S3, for the other metrics that were defined above. We note that for both the approaches yield high quality classification across all the metrics, and their performance is close to each other. While for this dataset, the DNN model performs somewhat better than the ensemble learning approach, it is important to acknowledge the high performance of the ensemble learning, as for larger datasets, it is known that the ensemble strategies can typically better mitigate the bias-variance trade-off.

Table S5

| Reference | Database | Data size | Classifiers | Features | AUC | Sensitivity | Specificity | Cohen's Kappa | Acc. |
|---|---|---|---|---|---|---|---|---|---|
| Our Model | DDH | 8227 | SVM,RF, DNN | 2D descr-iptors | 0.994 | 0.9436 | 0.9755 | 0.9195 | 0.959 |
| Czodrowski P, 2013 | ChemBL | 11958 | RF | RDKITdescrip-tors | 0.564 | 0.029-0.243 | - | - | 0.907 |
| Wang S et al, 2016 | - | 587 | Naïve Bayes, SVM | Pharmaco-phore hypothesis | 0.899 | 0.943 | 0.596 | - | 0.782-0.836 |
| Ogura et al, 2019 | hERG-Integrated Dataset | 291219 | SVM | 2D 3D descriptors, ECFP-4 structural fingerprints, PipelinePilot descriptors | 0.966 | 0.715 | 0.933 | 0.733 | 0.98 |
| Schyman P., 2016 | NationalCancer Insti. Database | 25000 | - | Accelrys extended connectivity fingerprints, conformations | - | 0.69 | 0.95 | - | 0.79 |
| Doddareddy et al | Dubus203, Literature368, Thai313 datasets | 7360 | LDA, SVM | Extended connectivity fingerprints, functional class fingerprints | 0.94 | - | - | - | 0.91 |
| Kwang-Eun et al | Pipeline Pilot (PP),FCFP 2, FCFP 4 and FCFP 6, R package | 5252 | DNN, NB, SVM,RF, Bagging | integer and binary type fingerprints | 0.95 | 0.626 | 0.986 | - | - |

| Chuipu Cai et al | ChEMBL, hERG K+ channel binding affinity, radioligand binding measurements on mammalian and non-mammalian cell lines, literature-derived data | 7889 | DNN, GCNN | Molecular Operating Environment descriptors, Mol2vec descriptors | 0.97 | 0.912 | 0.817 | - | 0.93 |

Table S5 provides an overall summary of the various existing methods for the task of classification of pharmaceutical compounds based on their hERG inhibition activity. We note that while there have been a few (but not many) methods to address this task, these involve different datasets, and different features. Thus, while such a summary is not a comparison, it does provide, under one roof, a perspective on data, methods, and can help in identifying scope of improvements in this area. In the table, some cells are blank as not all metrics are provided for all methods. We note that the methods do not yield a high performance across all the metrics. Some methods employ relatively small data. The work by Ogura et al, involves the largest data, but yields a low sensitivity and kappa coefficient. Such discrepancy suggests an issue with the data imbalance. Importantly, most of the approaches use various different features in their methodology. In contrast our approach uses only 2D features and yields a high performance across all metrics. A limitation of this work is that it also involves relatively less data, which we plan to address in future.

# References

1. Anderson, E., G.D. Veith, and D. Weininger. 1987. SMILES: A line notation and computerized interpreter for chemical structures. Report No. EPA/600/M-87/021. U.S. Environmental Protection Agency, Environmental Research Laboratory-Duluth, Duluth, MN 55804
2. B. Mehlig, 2019, Artificial Neural Networks. *arxiv* 1901.05639
3. Chuipu Cai, Pengfei Guo, Yadi Zhou, Jingwei Zhou, Qi Wang, Fengxue Zhang, Jiansong Fang, and Feixiong Cheng, 2015. Deep Learning-based Prediction of Drug-induced Cardiotoxicity. J Chem Inf Model. 2019 Mar 25; 59(3): 1073–1084.
4. Czodrowski, P., 2013 HERG Me Out. J. Chem. Inf. Model.53, 2240–2251 (2013).
5. David C. Kombo, Kartik Tallapragada, Rachit Jain, Joseph Chewning, Anatoly A. Mazurov, Jason D. Speake,
6. Terry A. Hauser, and Steve Toler,2012. 3D Molecular Descriptors Important for Clinical Success. J. Chem. Inf. Model. 2013, 53, 2, 327–342
7. Drug Discovery Hackathon, https://innovateindia.mygov. in/drug-ps/track-2-general-drug-discovery-includingcovid/ddt2-13/
8. Doddareddy MR; Klaasse EC; Ijzerman AP; 2010. Bender A Prospective Validation of a Comprehensive in Silico hERG Model and its Applications to Commercial Compound and Drug Databases. Chemmedchem 2010, 5, 716–729 [PubMed: 20349498].
9. Guillaume Lemaˆıtre and Fernando Nogueira and Christos K. Aridas,2017. Imbalanced-learn: A Python Toolbox to Tackle the Curse of Imbalanced Datasets in Machine Learning *Journal of Machine Learning Research.*
10. "hERG safety". Cyprotex, 9 October 2018. https://www.cyprotex.com/toxicology/cardiotoxicity/hergsafety
11. J. Cohen (1960). "A coefficient of agreement for nominal scales". Educational and Psychological Measurement 20(1):37-46.
12. Kola, I. & Landis, J.,2004. Can the Pharmaceutical Industry Reduce Attrition Rates? Nat. Rev. Drug Discov. 3, 711–715 (2004)
13. Laura E. Raileanu, Kilian Stoffel, 2004. Theoretical Comparison between the Gini Index and Information Gain Criteria. Annals of Mathematics and Artificial Intelligence 41, 77–93 (2004)
14. Leo Breiman, 2001. Random Forests, Statistics Department, University of California, Berkeley, CA 94720. *Springer*
15. Marchese Robinson, R. L.; Glen, R. C.; Mitchell, J. B., 2011. Development and comparison of hERG blocker classifiers: Assessment on different datasets yields markedly different results. *Mol. Inform. 2011,* 30, 443–458.
16. Matthews, B. W. (1975). Comparison of the predicted and observed secondary structure of T4 phage lysozyme. Biochimica et Biophysica Acta (BBA) - Protein Structure. 405 (2): 442–451.
17. Moriwaki H, Tian Y-S, Kawashita N, Takagi T (2018). Mordred: a molecular descriptor calculator. *Journal of Cheminformatics 10:4*.
18. N. V. Chawla, K. W. Bowyer, L. O. Hall, W. P. Kegelmeyer., 2011. SMOTE: Synthetic Minority Over-sampling Technique. arxiv - 1106.1813
19. Ogura, K., Sato, T., Yuki, H. et al.,2019. Support Vector Machine model for hERG inhibitory activities based on the integrated hERG database using descriptor selection by NSGA-II. Sci Rep 9, 12220 (2019).
20. Parasuraman, S.,2011. Toxicological screening." Journal of pharmacology & pharmacotherapeutics vol. 2,2 (2011): 74-9.
21. Peruri Venkata Anusha et. al., 2019. Detecting Outliers in High Dimensional Data Sets Using Z-Score Methodology. *Int. Journal of Innovative Tech. and Exploring Engg.*
22. RDKit. Open-source Chemiformatics http://www.rdkit.org.
23. Schyman, P., Liu, R. & Wallqvist, A.,2016. General Purpose 2D and 3D Similarity Approach to Identify hERG Blockers. J. Chem. Inf. Model. 56, 213–222 (2016)

24. Spiegel, Murray R.; Stephens, Larry J. (2008). Schaum's Outlines Statistics (Fourth ed.), McGraw Hill.
25. Dirk J. Snyders, 1999. Structure and Function of Cardiac Potassium Channels. *Cardiovasc. Res 42*, 377–390 (1999)
26. Wang, S. et al., 2012, ADMET Evaluation in Drug Discovery.12. Development of Binary Classification Models forPrediction of hERG Potassium Channel Blockage. *Mol.Pharmaceutics 9*, 996–1010 (2012).
27. Wang, S. et al.,2016. ADMET Evaluation in Drug Discovery. 16. Predicting hERG Blockers by Combining Multiple Pharmacophores and Machine Learning Approaches. Mol. Pharmaceutics 13, 2855–2866 (2016).
28. Yoon, Hye & Balupuri, Anand & Choi, Kwang-Eun & Kang, Nam (2020). Small Molecule Inhibitors of DYRK1A Identified by Computational and Experimental Approaches. International journal of molecular sciences. 21. 10.3390/ijms21186826.
29. Yongli Zhang, 2012. Qinggong College, Hebei United University, Tangshan, China. Support Vector Machine Classification Algorithm and Its Application.Information Computing and Applications. ICICA 2012. Communications in Computer and Information Science, vol 308. *Springer, Berlin, Heidelberg.*